\documentclass{bmvc2k}

\usepackage{amsmath}
\usepackage{amssymb}
\usepackage{mathtools}
\usepackage{amsthm}

\usepackage{algorithm}
\usepackage{algorithmicx}
\usepackage{algpseudocode}

\usepackage{comment}
\usepackage{tabularx}
\usepackage{framed,multirow,array}

\usepackage{microtype}
\usepackage{graphicx}
\usepackage{subfigure}
\usepackage{booktabs} 

\title{Segmentation of Prostate Tumour Volumes from PET Images is a Different Ball Game}

\addauthor{Shrajan Bhandary}{shrajan.bhandary@tuwien.ac.at}{1}
\addauthor{Dejan Kuhn}{dejan.kostyszyn@uniklinik-freiburg.de}{2}
\addauthor{Zahra Babaiee}{zahra.babaiee@tuwien.ac.at}{1}
\addauthor{Tobias Fechter}{tobias.fechter@uniklinik-freiburg.de}{2}
\addauthor{Simon K.B. Spohn}{simon.spohn@uniklinik-freiburg.de}{2}
\addauthor{Constantinos Zamboglou}{constantinos.zamboglou@uniklinik-freiburg.de}{2}
\addauthor{Anca-Ligia Grosu}{anca.grosu@uniklinik-freiburg.de}{2}
\addauthor{Radu Grosu}{radu.grosu@tuwien.ac.at}{1}

\addinstitution{
 Technische Universität Wien, Vienna, Austria
}
\addinstitution{
 Department of Radiation Oncology, University Medical Centre Freiburg, Germany
}

\runninghead{Authors}{Segmentation of Prostate Tumours from PET Images}

\begin{document}

\maketitle

\begin{abstract}
Accurate segmentation of prostate tumours from PET images presents a formidable challenge in medical image analysis. Despite considerable work and improvement in delineating organs from CT and MR modalities, the existing standards do not transfer well and produce quality results in PET related tasks. Particularly, contemporary methods fail to accurately consider the intensity-based scaling applied by the physicians during manual annotation of tumour contours. In this paper, we observe that the prostate-localised uptake threshold ranges are beneficial for suppressing outliers. Therefore, we utilize the intensity threshold values, to implement a new custom-feature-clipping normalisation technique. We evaluate multiple, established U-Net variants under different normalisation schemes, using the nnU-Net framework. All models were trained and tested on multiple datasets, obtained with two radioactive tracers: [$^{68}$Ga]Ga-PSMA-11 and [$^{18}$F]PSMA-1007. Our results show that the U-Net models achieve much better performance when the PET scans are preprocessed with our novel clipping technique.     

\end{abstract}

\begin{figure*}[t]
\centering
\includegraphics[width=\linewidth]{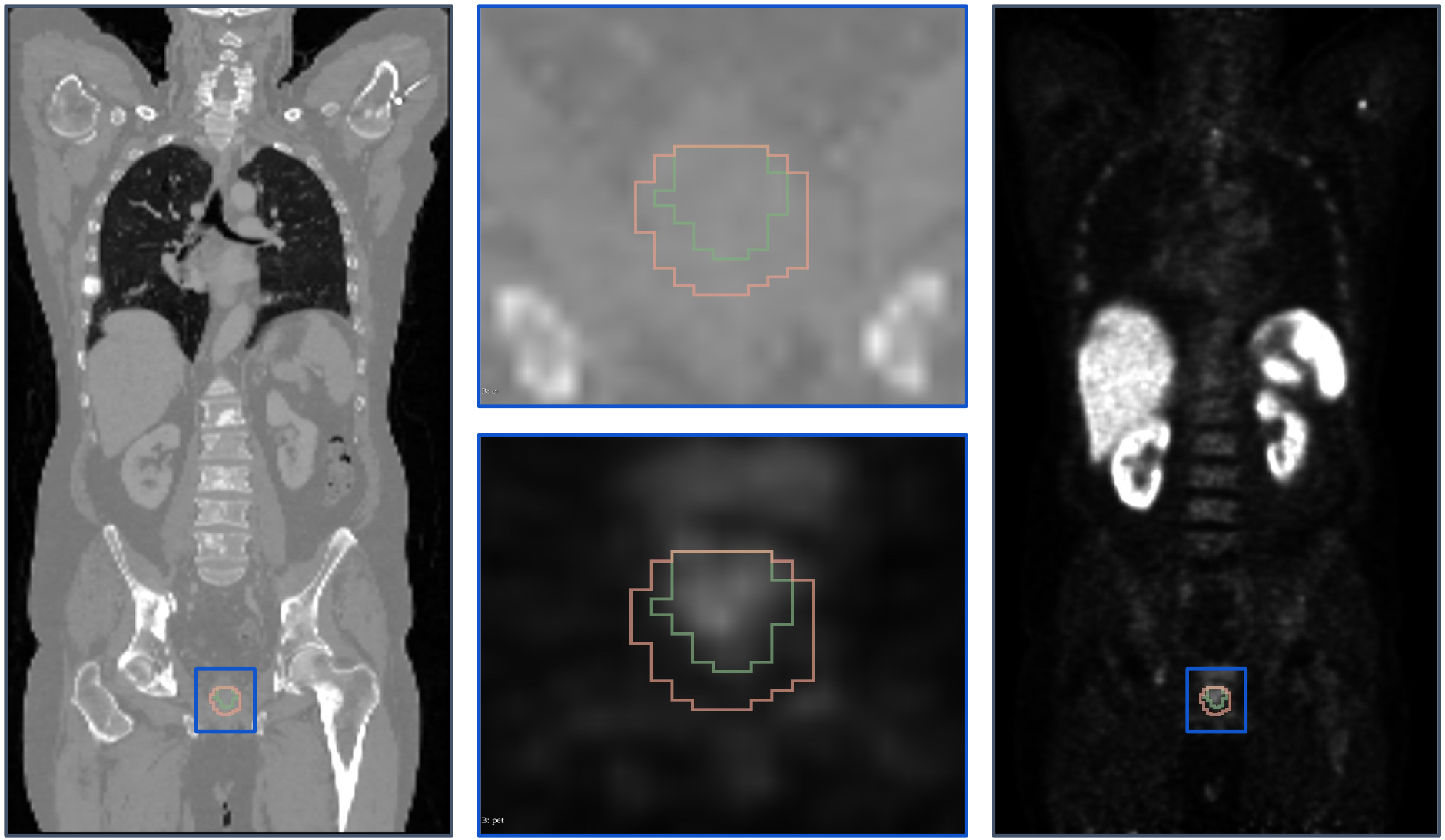}
\vspace{-6mm}
\caption{Example of CT (left) and [$^{68}$Ga]Ga-PSMA-11 PET (right) images with annotated prostate gland (red) and tumour (green). Center: Cropped pelvic regions with the prostate and tumour (top: CT and bottom: PET). These images highlight the importance of correct modality for RT, as the tumour volume is more pronounced in PET than in CT.}
\label{fig:ct_pet_labels}
\vspace{-3mm}
\end{figure*}

\begin{algorithm}[t]
\caption{FCN Algorithm for GTV Segmentation. The algorithm finds the optimal upper threshold limit based on the SUV values of all PET images in the training dataset.}
\label{alg:semi_contouring}
\begin{algorithmic}
\vspace*{1mm}
\State {\bfseries Input:} data - $x$, labels - $y$
\State {\bfseries Output:} maximum threshold limit - $maxT$
\vspace*{1mm}
\For{$p=20\%$ {\bfseries to} $70\%$ {\bfseries increment} $2\%$}
    \For{$i=1$ {\bfseries to} \textit{sample-count}($x$)}
    \State \textit{threshold} $= p~*$ \textit{max}$(x_i) * 0.01$ \Comment{Save \textit{threshold} value}
        \For{$j=1$ {\bfseries to} \textit{voxel-count}$( x_i)$}
            \If{$x_i(j) \geq $ \textit{threshold}} 
            \State $y'_i(j) = 1$ 
            \Else
            \State $y'_i(j) = 0$
            \EndIf
        \EndFor
        \State Calculate \textit{DSC}$(y'_i,y_i)$ and \textit{NSD}$(y'_i,y_i)$ \Comment{Save \textit{DSC} and \textit{NSD} metric results}

    \EndFor
    \State Calculate average \textit{DSC}$(y',y)$ and average \textit{NSD}$(y',y)$ \Comment{for each \textit{p}}
\EndFor
\State Find $p_{max}DSC$ where, $p_{max}=p$ for highest average \textit{DSC} 
\State Find $p_{max}NSD$ where, $p_{max}=p$ for highest average \textit{NSD} 

\State Find average SUV \textit{threshold} ($t_{max}DSC$) value at $p_{max}DSC$ 
\State Find average SUV \textit{threshold} ($t_{max}NSD$) value at $p_{max}NSD$ 

\State {\bfseries return} \textit{maxT} =  average of $t_{max}DSC$ and $t_{max}NSD$


\end{algorithmic}
\end{algorithm}
\vspace*{-6mm}

\begin{figure*}[t]
\centering
\includegraphics[width=\linewidth]{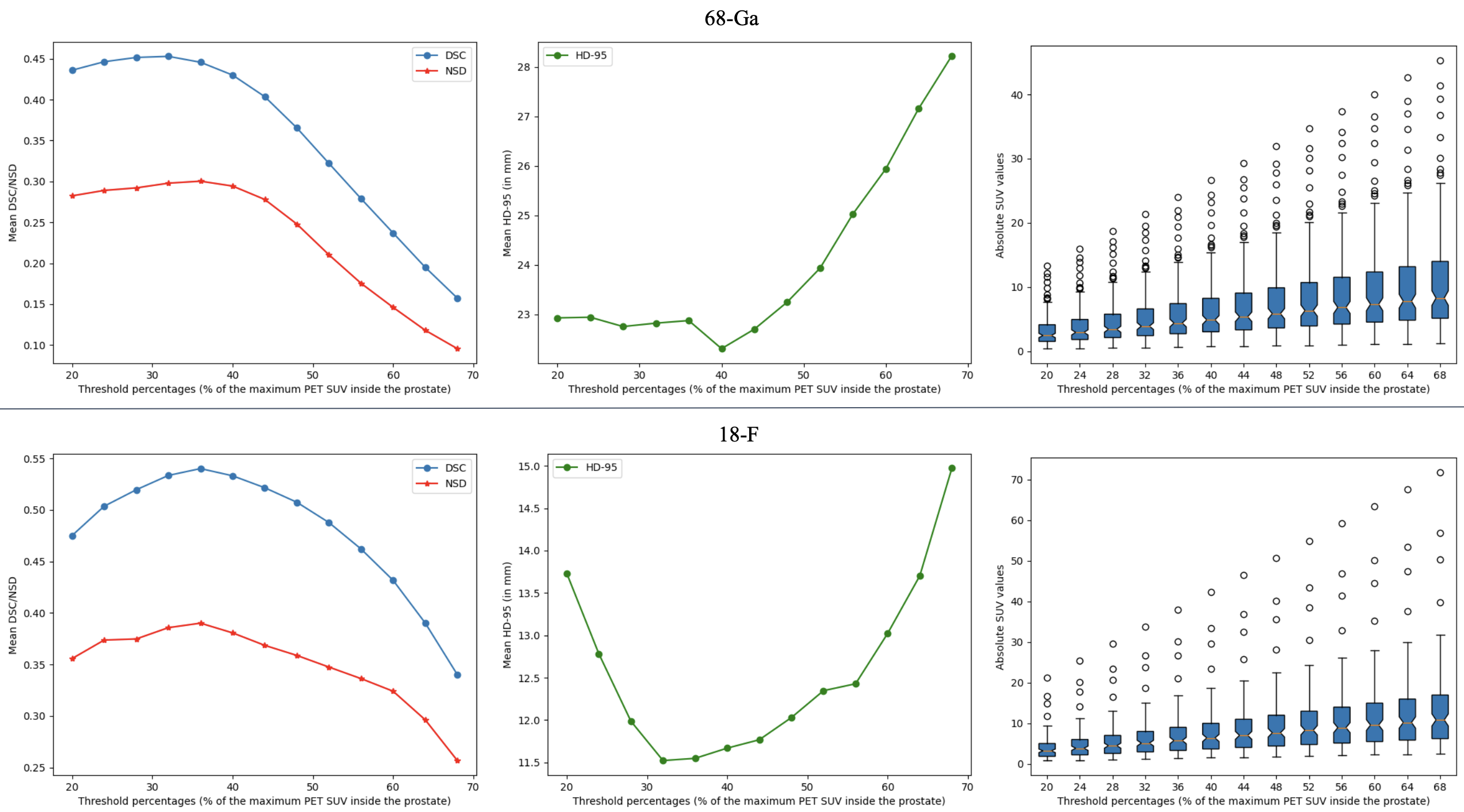}
\caption{Left and Center: Performance of semi-automatic contouring approaches on the intra-prostatic SUV, of the PET images for 68-Ga and 18-F tracers. All voxels with uptake values equal-to and above the SUVmax$\%$ threshold, are considered as tumour, and the rest as background. For Dice coefficient (DSC) and normalised surface dice (NSD), the higher their value the better the prediction, whereas, for the Hausdorff Distance 95\% percentile (HD-95) is the other way around. 
The best performance for all three metrics is achieved between 30\%-40\%.
Right: The absolute threshold uptake values of all images for a given percent. Most of the average threshold values lie in the range 3-10.}
\label{fig:ThresholdPlots}
\end{figure*}

\section{Introduction}
Prostate Cancer (PCa) is an ubiquitous malignancy in men that accounts for nearly 30\% of all diagnosed cancers in the USA and Europe~\cite{sung2021global, MARHOLD2022304, siegel_2024_CancerStatistics}. 
Radiotherapy (RT) is one of the primary treatment approaches, that demands precise localization of gross-tumour volumes (GTV)   and organs at risk (OAR)~\cite{kerkmeijer2021focal, zamboglou2022psma}. 
Common medical imaging modalities, such as computed tomography (CT) and magnetic resonance imaging (MR) provide detailed anatomical information. However, GTV segmentation from CT and MR volumes is challenging~\cite{eiber2016simultaneous,spohn2020comparison}.

Recently, positron-emission tomography (PET) scans demonstrated a significant potential for identifying tumour volumes, particularly intra-prostatic lesions using prostate-specific membrane antigen (PSMA). 
PET images excel in characterizing metabolic activity using the standardized uptake value (SUV), which is directly proportional to the intensity of radioactive tracer uptake by cancer cells. This phenomenon is evident in the PET scans in Figure~\ref{fig:ct_pet_labels}.

To remedy the challenges posed by manual segmentation~\cite{draulans202168ga,spohn2020comparison}, 
deep-learning-based (DL) approaches, such as U-Nets~\cite{ronneberger_u_net_2015, cicek_3d_2016}, have been investigated to automatically segment 
PCa from PET scans~\cite{kostyszyn_intraprostatic_2020, xu_automatic_2023, Holzschuh2023}. In spite of these works, research on tumour delineation from PET images is relatively limited compared to other clinical modalities. Hence, there is a scarcity of well-established configurations and preprocessing pipelines for PET segmentation tasks. Current research works rely upon normalisation schemes tailored for CT and MR volumes~\cite{kostyszyn_intraprostatic_2020, xu_automatic_2023, Holzschuh2023,isensee2023look, murugesan2023autopet, hadlich2023autopet}. These preprocessing techniques may not be suitable for PET images, since they factor SUV values from the full image. Whereas, PET scans illustrate intensity values based on the localised and concentrated metabolic activities of the body region~\cite{Zamboglou2019, spohn2020comparison, zamboglou2022psma}. Since the network performance is reliant on normalisation methods~\cite{litjens_2017, nnU_Net_2020}, there is an urgent need for robust preprocessing steps in PET segmentation.

In this work, we overcome the shortcomings of the existing research in two steps. First, we investigated the various normalisation techniques currently in use, and scrutinised their impact on the performance of GTV segmentation from PET images. Additionally, we leveraged the insights of the physicians during manual annotation, by incorporating SUV-threshold values in our preprocessing steps. This novel normalization method is called as feature-clipping normalisation (FCN). 
The FCN algorithm helps to find all-encompassing optimum threshold value for different datasets, especially when dealing with a specific type of PSMA-tracer.

In the second step, to assess the benefits of our new FCN method, we compared four U-Net variants: Classic U-Net, Attention-U-Net, and their inductive-bias extensions, IB-U-Net, and IB-Attention-U-Net, respectively. We conducted our experiments with multiple PCa datasets with different tracers, namely [$^{68}$Ga]Ga-PSMA-11 (68-Ga) and [$^{18}$F]PSMA-1007 (18-F), and objectively evaluated the models.
The results showed that our SUV threshold-based normalisation improves the accuracy of all the networks.
%
In summary:
\vspace{-2mm}
\begin{itemize}
\setlength\itemsep{-1mm}
\item\textit{Our main contribution in this paper, is to develop a novel feature-clipping normalisation method, we call FCN, that provides custom clipping limits according to the threshold values used by physicians while performing the manual segmentation.}

\item\textit{Another contribution in this paper is to empirically demonstrate that our new feature-clipping normalisation improves the accuracy of the networks for PCa segmentation from PET images, irrespective of the PSMA tracer and SUV-threshold values.}
\item\textit{Finally, we also implemented our FCN method and all the U-Net variants used (U-Net, IB-U-Net, Attention U-Net and IB-Attention U-Net) in the popular state-of-the-art nn-Unet framework. The code is open source, and the framework is easy to use.}
\end{itemize}

\section{Literature Review}
\label{section:Literature_Review}

\subsection{U-Net variants}
Since the release of the 2D U-Net architecture in 2015~\cite{ronneberger_u_net_2015}, numerous revisions and variants have been proposed~\cite{Punn2022}. Many of these variations propose extensions and advances, and have achieved state-of-the-art (SOTA) performances: V-Net~\cite{vnet_2016}, Attention U-Net~\citep{attention_unet_2018}, U-Net++~\citep{unet_pp_2018}, and SegResNet~\citep{segresnet2018}.~\citet{bhandary2022ib} compared the performances of these models using the nn-UNet framework~\cite{nnU_Net_2020} on multiple medical segmentation tasks. Inspired by the inner workings of the vertebrate retina, the authors also introduced inductive biased (IB) On-and-Off convolutional filters in the second encoder of the U-Net architecture. The addition of IB filters in the network structure improved the accuracy of segmentation networks, and made them robust against artefacts. They demonstrated superior performance for small-sized datasets, especially in tasks pertaining to the prostate gland~\cite{bhandary2022ib}.

\subsection{GTV segmentation}
Before we discuss various DL-based algorithms, we first evaluate threshold-based PET image segmentation approaches.~\citet{Zamboglou2019} examined $20$ patients with PCa who underwent 68-Ga PSMA-11-PET/CT followed by radical
prostatectomy. They carried out manual and semi-automatic segmentations, and concluded that SUV capping
($SUV$min-max $0$–$5$) or thresholding ($20\%$ of $SUV$max) could provide high sensitivity, and should be considered for
PSMA-PET-based focal therapy approaches. The study by ~\citet{spohn2020comparison} subjected $10$ patients to 18-F PSMA-1007 PET, and then performed radical prostatectomy. They observed that manual contouring with PET scaling of $SUV$min-max $0$–$10$, or a semi-automatic approach with a threshold of $20\%$ of $SUV$max offer the best results.

~\citet{TAMAL2020e05267} conducted a review of different fixed and adaptive threshold-based PET image segmentation approaches under a common mathematical framework. The author highlighted the advantages and disadvantages of the threshold-based methods from the perspectives of diagnosis, treatment planning, and response assessment. In their experiments, they also observed that a fixed threshold-based method is dependent on the tumour, to background ratio, and the size of the tumour, and therefore, recommended adaptive threshold-based methods.~\citet{TAMAL2020e05267} advocated that SUV-based threshold methods hold merit, and proposed that advanced adaptive approaches such as DL algorithms could improve GTV segmentation. 

Over the past decade, convolutional neural networks (CNNs), such as the U-Nets, have demonstrated their ability as an invaluable tool that can adapt automatically to complex medical segmentation tasks. However, depending upon image modality, segmentation of the prostate gland and the associated tumour volumes is extremely challenging. For example, some of the works that performed PCa segmentation from MR images did not achieve good results~\cite{keno_bressem_2022_6481141, saha_anindo_2022_6522364}. On the other hand, there are some instances where U-Net based GTV segmentation have been moderately successful.


%
One of the earliest works on prostate tumour delineation from PET images by~\citet{kostyszyn_intraprostatic_2020}, employed a 3D U-Net~\cite{cicek_3d_2016}. The dataset consisted of 68-Ga PSMA-11-PET images, and were normalised with global intensity values. Likewise,~\citet{Holzschuh2023} trained a 3D U-Net on a 18F-PSMA-PET dataset, and evaluated the resultant model on internal and external test sets. 
In the preprocessing steps, although the authors cropped the PET images to the pelvic region, an arbitrary $SUV$max value of $15$ was used to clip the data. The overall segmentation performances in~\citet{kostyszyn_intraprostatic_2020} and~\citet{Holzschuh2023} were good. However, in both instances, we also observed a lack of meaningful comparison with other networks, and did not find a rationale behind choosing the normalisation methods.

Recently, AutoPET-II challenge~\cite{gatidis_2023_autopet} was organized with the goal to segment cancer lesions, such as malignant melanoma, lymphoma, and lung cancer. 
In spite of large and multifaceted network architectures, the segmentation accuracies of the top ranking submissions were low. A possible explanation could be that the participants in the competition utilised various normalisation schemes, normally applied to CT and MR images~\cite{isensee2023look, murugesan2023autopet, hadlich2023autopet}. 


These studies highlight the rapid advancements in the field of PET imaging, particularly PSMA-PET, for prostate cancer. Although the integration of deep learning networks has enabled a more accurate and efficient segmentation of GTV, there is still considerable room for improvement, and standardisation.

\section{Methods and Materials}
\label{sec:methods-and-materials}

\subsection{Datasets}
\label{sec:datasets}
In this paper, we used two in-house prostate-specific membrane antigen (PSMA) PET datasets. Simultaneously, CT scans were also captured to locate OARs. The datasets comprised of PET images that were collected using two unique tracers: Gallium (68-Ga) gozetotide or Gallium (68-Ga) PSMA-11, and piflufolastat (18-F). 

The multi-institutional 68-Ga dataset consists of a combined total of 168 scans collected at Centres A (n = 142), and B (n = 26). Out of the 168 volumes, 151 collated from Centres A (n = 125) and B (n = 26), respectively, were used during training. The remaining 17 volumes from Centre A were used for final testing. The second, 18-F PSMA-PET data consists of 131 training images from Centre A. Testing was done on 50 patient images that were collected from an independent internal cohort from Centre A (n = 19), and from an independent external cohorts obtained from Centre C (n = 14). 

In both datasets, the manual segmentation of the prostate on CT, and the tumour volumes on the PET images were generated by expert physicians of the respective groups in consensus. To decrease the inter-observer variability, validated techniques were used to annotate the tumour on PET. For the 68-Ga PSMA-PET dataset, a scaling range of $SUV$min$-SUV$max of [$0-5$] was used by the physicians during annotations, whereas, $SUV$min$-SUV$max of [$0-10$], was used for the 18-F dataset. The scaling was applied uniformly cross all the images based on the type of tracer, including in-house and external sets.


All the images (PET volume, prostate contour and PCa ground-truth labels) were resampled to a voxel size of $2 \times 2 \times 2$ $mm^3$. The PET scans were resampled with B-spline interpolation, while the annotated prostate and tumour contours were resampled using nearest neighbour interpolation technique. The PET volumes and the GTV ground-truth labels were cropped around the prostate, and we added a second channel, consisting of the prostate mask, along with the PET image as input. This improved overall segmentation performance. Finally, all of these volumes were converted to the nnU-Net framework format for the experiments.

\subsection{Existing Normalisation Schemes}
\label{sec:existing_norm_schemes}
One of the crucial aspect of this study is the evaluation of different normalisation schemes on PET-based tumour segmentation performance. The nnU-Net framework, by default, offers two of the most popular normalisation techniques, namely $Z$-score for MR and other non-CT images, and~\textit{global normalisation with percentile clipping} for CT images. Equation~\eqref{eq:z_score} gives the formula for computing the $Z$-score using local (single image) parameters:
\begin{equation}
\label{eq:z_score}
\begin{array}{l@{\ }l}
x'_i(j) &= \frac{x_i(j) - \mu_i}{\sigma_i}, \qquad x_i \in x = {x_0, ..., x_N-1}
\end{array}
\end{equation}
where, $x$ is the full training set with $N$ samples, $x_i$, the $i$th sample, $x_i(j)$, the raw value at voxel $j$, $\mu_i$, the mean of all voxels in sample $i$, $\sigma_i$, the standard deviation of all voxels in sample $i$, and $x'_i$, the standardized value of the $i$th sample. 

Generally, for CT images, each voxel intensity of an image is first clipped to a minimum of $0.5$ percentile and a maximum of $99.5$, of all voxels in the entire training dataset, as given by Equation~\eqref{eq:clip}. Then Equation~\eqref{eq:ct} is used to standardize the clipped voxel value, using the global parameters (across all images). Thus:

\begin{equation}
\label{eq:clip}
\begin{array}{l@{\ }l}
x(J)_\text{0.5percentile}~ \leq x_i(J)~ \leq x(J)_\text{99.5percentile}
\end{array}
\end{equation}
\vspace*{-4mm}\begin{equation}
\label{eq:ct}
\begin{array}{l@{\ }l}
x'_i(j) &= \frac{x_i(j) - \mu_J}{\sigma_J}, \qquad x_i \in x = {x_0, ..., x_N-1} 
\end{array}
\end{equation}
where, $x$ is the full training set with $N (0-n)$ samples, $J$, represents all the voxels in all images $x$, $x_i$ is the $i$th sample, $x_i(j)$ is the raw value at voxel $j$, $\mu_J$, the mean of all voxels in $x$, $\sigma_J$, the standard deviation of all voxels in $x$, and $x'_i$ is the standardized value of the $i$th sample. 

In addition to $Z$-score and CT(~\textit{global normalisation with clipping}), we added a clipping-based normalisation, based on the uptake threshold limits applied during manual delineation in~\citet{Holzschuh2023}. We call this \textit{fixed clip}, and the minimum threshold (min$T$) was $0$, whereas, the maximum limit (max$T$) was set to $15$~\cite{Holzschuh2023}. However, it should be noted that normalisations were applied to only the first channel consisting of PET images, as the second is a binary mask of the prostate contours.

\subsection{Feature Clipping Normalisation (FCN)}
\label{sec:fcn}
The motivation and idea behind the algorithm is based on the combination of the manual and semi-automatic contouring technique recommended by the physicians in~\citet{Zamboglou2019} and~\citet{spohn2020comparison}, and the review by~\citet{TAMAL2020e05267}. As mentioned in Section~\ref{sec:datasets}, both 68-Ga and 18-F PSMA-PET datasets were manually annotated after scaling them to a particular fixed value ($5$ and $10$ for 68-Ga and 18-F, respectively). Instead of using these fixed maximum bounds to scale the intensity values, our FCN method automatically finds the upper clipping value (max$T$) for preprocessing a specific dataset. Since the least voxel SUV is for most PET images is usually $0$, the lower clipping value (min$T$) was set to $0$. 

The method to find the optimal upper bound is illustrated in Algorithm~\ref{alg:semi_contouring}. We decided to use a search range from 20\% to 70\% of $SUV$max with an increment of 2. As mentioned in Algorithm~\ref{alg:semi_contouring}, if the SUV for a voxel was greater than the threshold value, then it was considered as a tumour, otherwise normal. Using this, a prediction image was obtained and compared against the ground-truth label using NSD ($\uparrow$), HD-95 ($\downarrow$) and DSC ($\uparrow$) metrics. The evaluation results between the ground-truth labels, and predictions using the metrics, are showcased in Figure~\ref{fig:ThresholdPlots} (left and centre). The percentage-wise absolute threshold ($t$) value from Algorithm\ref{alg:semi_contouring} for each image is displayed as a box plot (right) in Figure~\ref{fig:ThresholdPlots}. 

As shown in the metrics graphs of Figure~\ref{fig:ThresholdPlots}~(left and centre), our FCN semi-automatic contouring method achieved the highest performance, for percentages between $30$ to $40$. Above $40\%$, the accuracies dip, and this is true for both datasets. The top \textit{NSD} value is $0.30$ for the 68-Ga tracer, and $0.38$ for the 18-F tracer. Similarly, the top \textit{DSC} value is $0.45$ for the 68-Ga tracer, and $0.54$ for the 18-F tracer. From the box plots of Figure~\ref{fig:ThresholdPlots}, the average SUV-threshold for the percent range $30$ to $40$ is between $3$ and $10$ for both the 68-Ga and 18-F datasets. The final max$T=5.142$ for 68-Ga, and max$T=8.736$ for 18-F training datasets. This is approximately equal to the SUV limits used during manual segmentation. We would like to point that, we did not consider the results of HD-95 metric to determine the upper threshold limit (max$T$). This is because, hausdorff distance does not have an upper bound (range: $0-\infty$), and therefore, gives inaccurate measurements for empty volumes~\cite{maier2022metrics}. 

\subsection{Implementation of our framework}
We used the versatile and robust data pipeline of the nnU-Net framework (version $2$)~\cite{nnU_Net_2020} to build our tumour-segmentation solution. As mentioned in Section~\ref{sec:existing_norm_schemes}, $Z$-score and CT(~\textit{global normalisation with percentile clipping}) are already available in the nnU-Net framework. The~\textit{fixed clip} and FCN normalisation methods implemented by us can be utilised by specifying the type of normalisation scheme (when required) in the~\textit{dataset.json} file. For a given PSMA-PET dataset, the upper bound (max$T$) is calculated during the~\textit{experiment planning} and ~\textit{dataset fingerprint extraction} phases. Finally, all the PET images are clipped to min$T = 0$ and the max$T$ value obtained previously. We evaluated four different 3D models (did not use 2D versions), namely the U-Net, IB-U-Net, Attention-U-Net, and IB-Attention-U-Net. The IB-extended versions were chosen for their robustness abilities against distribution shifts~\cite{bhandary2022ib}. The new normalisation schemes (including the FCN) and the additional network architectures were implemented in PyTorch~\cite{pytorch_2019_NIPS}, in accordance to the nnU-Net framework~\cite{nnU_Net_2020} guidelines. 

The nnU-net framework has data pipelines that carry out various data augmentations. For both 68-Ga and 18-F datasets, a mini-batch size of 6 was used. In addition to the final prediction, the outputs of deep supervision layers were used for the final loss calculation. We used a compound loss function, which is a combination of cross-entropy loss and dice loss~\cite{vnet_2016}, to optimise the networks. All the models were trained for 1000 epochs, and a sliding window technique with an overlap size of $50\%$ was used during inference. Due to the possibility of multiple tumour lesions in a given image, the default post-processing step (retain only the largest connected component) was not applied.

\subsection{Experiments and Results}
We conducted our GTV segmentation experiments in two ways with U-Nets using the nnU-Net framework. The first procedure used existing normalisation schemes ($Z$-score, CT and fixed clip); whereas, the second method used the automatic FCN mentioned in Section~\ref{sec:fcn}. For both stages, we investigated the different U-Net variants using 5-fold cross-validation (CV) experiments. Each model was trained on both the 68-Ga and 18-F PSMA-PET datasets with four different normalisation schemes. The experiments were run on an NVIDIA Titan RTX with 24 GB memory. We evaluated the overall performances using three metrics: NSD ($\uparrow$), HD-95 ($\downarrow$) and DSC ($\uparrow$), however, for brevity we only present the results using NSD in this paper. Segmentation performances using DSC and HD-95 are given in the appendix.


\begin{table*}[t]
\fontsize{9}{10}\selectfont
\centering
\begin{tabular}{|c|c|c|c|c|c|c|c|}
\hline                  
PSMA & Training & Testing & Model & Z-score & CT & fixed clip: & FCN \\
tracer & data & data & name &~\cite{nnU_Net_2020} &~\cite{nnU_Net_2020} &$0$-$15$~\cite{Holzschuh2023} & (ours) \\ 
\hline
\multirow{4}{*}{68-Ga} & & & UNet & $0.579$ & $0.609$ & $0.637$ & $\mathbf{0.661}$ \\
& Centres A + B & Centre A & IB-UNet & $0.616$ & $0.637$ & $0.668$ & $\mathbf{0.699}$ \\
& \textit{n} = 151 & \textit{n} = 17  & Att. UNet & $0.553$ & $0.650$ & $0.652$ & $\mathbf{0.666}$ \\
& & & IB-Att. UNet & $0.601$ & $0.664$ & $0.671$ & $\mathbf{0.705}$ \\
\hline
\multirow{8}{*}{18-F} &  & & UNet & $0.657$ & $0.713$ & $0.700$ & $\mathbf{0.738}$ \\
& & Centre A & IB-UNet & $0.686$ & $0.720$ & $0.718$ & $\mathbf{0.747}$ \\
& & \textit{n} = 19 & Att. UNet & $0.656$ & $0.726$ & $0.701$ & $\mathbf{0.741}$ \\
& Centre A  & & IB-Att. UNet & $0.677$ & $0.751$ & $0.685$ & $\mathbf{0.761}$ \\
\cline{3-8}
& \textit{n} = 131 & & UNet & $0.545$ & $0.587$ & $0.601$ & $\mathbf{0.644}$ \\
& & Centre C & IB-UNet & $0.598$ & $0.596$ & $0.616$ & $\mathbf{0.657}$ \\
& & \textit{n} = 14 & Att. UNet & $0.543$ & $0.605$ & $0.608$ & $\mathbf{0.653}$ \\
& & & IB-Att. UNet & $0.603$ & $0.600$ & $0.627$ & $\mathbf{0.667}$ \\
\hline
\end{tabular}
\vspace{2mm}
\caption{Comparison of U-Net variants all implemented by us in the same nnU-Net framework, for different normalisation methods, using NSD metric ($\uparrow$).~\textit{Z-score} standardises voxels using local mean and standard deviation.~\textit{CT} first does percentile clipping and then standardises voxels using global mean and standard deviation.~\textit{fixed clip} limits intensities to pre-defined lower and upper bounds. Statistical significance difference between best performing FCN preprocessing scheme against the rest of the normalisations is in bold.}
\label{table:AllNormResults_NSD_Metrics}
\end{table*}

\begin{figure*}[t]
\centering
\includegraphics[width=\linewidth]{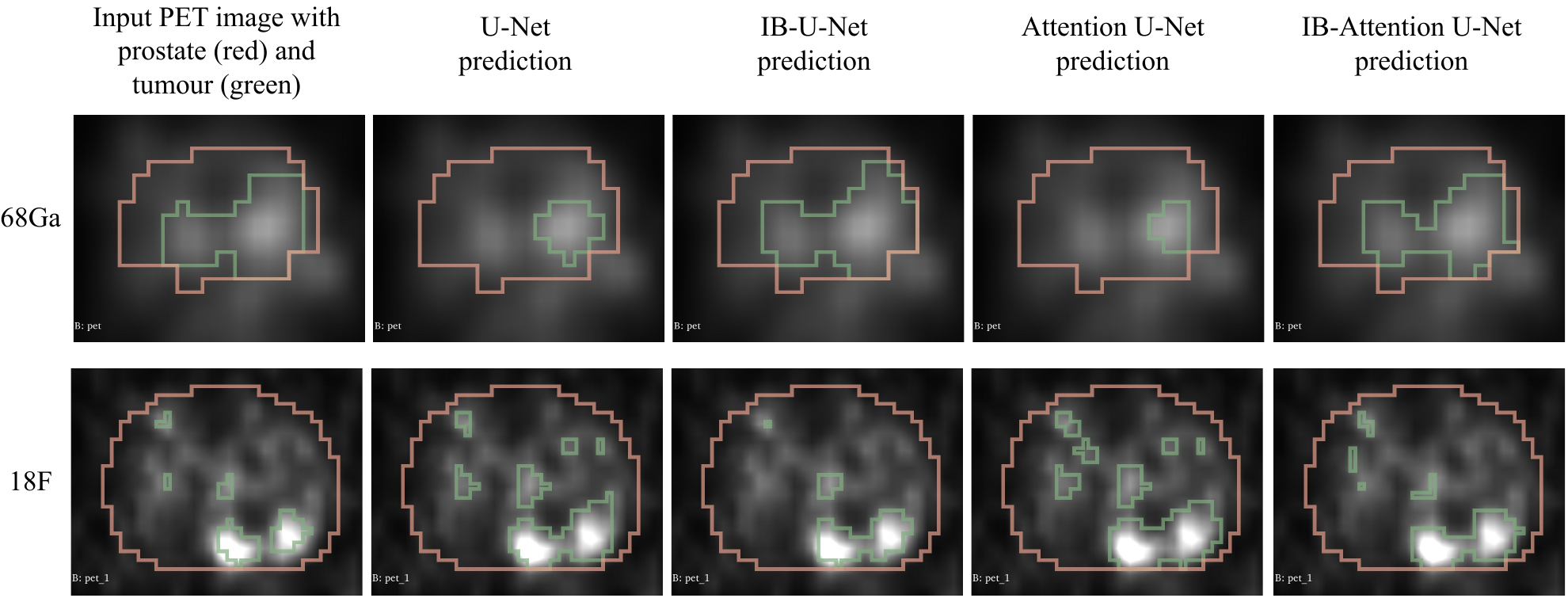}
\vspace*{-2mm}
\caption{A qualitative accuracy comparison of U-Net and Attention U-Net and their IB extended variants on the prostate tumour segmentation task. All the PET images were scaled using the FCN algorithm, and then trained using the U-Net models. As one can observe, the IB-versions perform better (fewer instances of false positives), with the IB-Attention-Net, achieving the best performance. Furthermore, the results show that the tumours are prominently differentiable from the background with 18-F tracer, in contrast to 68-Ga.}
\label{fig:SegmentationComparison}
\end{figure*}

Table~\ref{table:AllNormResults_NSD_Metrics} showcases the segmentation accuracy of all the four models for both datasets across four different normalisation methods. It is clearly evident that the networks that we preprocessed with FCN are superior to the existing normalisation schemes. We also conducted model-wise statistical significance tests (Wilcoxon-signed rank test) to compare the differences in performance between the top ranking normalisation scheme (FCN) versus the remaining three (Z-score, CT and SUV clip). This is highlighted in bold face in Table~\ref{table:AllNormResults_NSD_Metrics}, and please refer to the appendix for the detailed results. The low $p$ values ($p\,{<}\,0.05$) for the NSD metric in Table~\ref{table:AllNormResults_NSD_Metrics} indicate our feature-clipping technique is the best normalisation scheme for the both the tracers, 68-Ga and 18-F. 
For both tracers, the IB-variants have surpassed the original versions, and the highest segmentation accuracy is attained by the IB-Attention U-Net. 
Figure~\ref{fig:SegmentationComparison} shows the qualitative behaviour of the four models, for their best normalisation schemes, respectively. The rank of the normalisation methods in descending order is FCN, CT (68-Ga) or fixed clip (18-F), and $Z$-score.  

\section{Discussion and Conclusion}
\label{sec:discussion}

In this paper, we have successfully demonstrated the effectiveness and the practical applicability of U-Net-based CNN architectures in the segmentation of tumour volumes from PSMA-PET images. More importantly, we have shown that our FCN approach offers a significant advancement over semi-automatic delineation methods, and other diverse normalisation techniques. We have also investigated four different U-Net variants across two different PSMA tracers. Our results indicate that the FCN algorithm helps to
provide a more accurate, more efficient, and more reproducible means of tumour volume analysis.
%
The FCN algorithm was developed to encapsulate the intuition of the physicians and doctors to improve segmentation from PET images. By clipping the PET images, the networks are able to better focus of on the local information around the prostate. This advantage is not present in other normalisation techniques, such as $Z$-score and CT.

Despite these advancements, our study acknowledges certain limitations, such as the size and diversity of the dataset used. Our work is currently limited to PCa segmentation, that is localized to the pelvic region. None of the models achieved superior performance (NSD$\,{>}\,0.8$), and this is because like CT and MR scans, PET images are highly diffused. It is difficult to estimate the tumour growth, prostate whole-gland, and zonal boundaries with high certainty due to imaging limitations. Furthermore, the  high degree of inter-observer heterogeneity, during manual segmentation of tumours from PET images, could further exacerbate the difficulties of supervised learning approaches, such as the U-Net. Because of this, it is possible that the models start replicating the subjective errors of the annotators. 

On the positive note, prostate GTV segmentation from PET images is better when compared to CT and MR scans, as the SUV intensities are extremely beneficial in locating the lesions in the images. As shown in Figure~\ref{fig:SegmentationComparison}, the models are quite adept in segmenting the tumour volumes, prominently from PSMA-PET images for the 18-F tracer. In our future work, to ensure generalizability and robustness, we aim to validate the proposed normalisation techniques, and segmentation algorithms on larger and more heterogeneous datasets, such as the AutoPET-III~\cite{gatidis_2023_autopet}. Additionally, we will also explore the integration of these CNN-based techniques, into open-source software for medical image computing, such as the 3D Slicer software~\cite{slicer3dpaper}. Another avenue of future research, would be to take advantage of the trained model weights, and use them to classify the tumour grades based on Gleason Scores.

In conclusion, this paper demonstrates the effectiveness of U-Nets, and their inductive biased versions, in accurately delineating intra-prostatic GTV in PSMA-PET images. Our results show that by employing an appropriate normalisation technique, in conjunction with the labelling protocols used by the physicians, helps to improve the segmentation performance. Moreover, when compared to their seminal architectures, the IB-extended UNets are more versatile and robust in accurately delineating PCa from PET images.

\section*{Acknowledgement}
This research was funded in part by the Austrian Science Fund (FWF): I 4718, and the Federal Ministry of Education and Research (BMBF) Germany, under the frame of the Horizon-2020 ERA-PerMed call JTC-2019 Project Personalized Medicine: Multidisciplinary Research Towards Implementation (PersoRad). Z. B. was supported by the Doctoral College Resilient Embedded Systems, which is run jointly by the TU Wien's Faculty of Informatics and the UAS Technikum Wien.

\bibliography{main}

\begin{thebibliography}{32}
\providecommand{\natexlab}[1]{#1}
\providecommand{\url}[1]{\texttt{#1}}
\expandafter\ifx\csname urlstyle\endcsname\relax
  \providecommand{\doi}[1]{doi: #1}\else
  \providecommand{\doi}{doi: \begingroup \urlstyle{rm}\Url}\fi

\bibitem[Bhandary et~al.(2022)Bhandary, Babaiee, Kostyszyn, Fechter, Zamboglou,
  Grosu, and Grosu]{bhandary2022ib}
Shrajan Bhandary, Zahra Babaiee, Dejan Kostyszyn, Tobias Fechter, Constantinos
  Zamboglou, Anca-Ligia Grosu, and Radu Grosu.
\newblock Ib-u-nets: Improving medical image segmentation tasks with 3d
  inductive biased kernels.
\newblock \emph{arXiv preprint arXiv:2210.15949}, 2022.

\bibitem[Bressem et~al.(2022)Bressem, Adams, and
  Engel]{keno_bressem_2022_6481141}
Keno Bressem, Lisa Adams, and Günther Engel.
\newblock Prostate158 - training data, April 2022.
\newblock URL \url{https://doi.org/10.5281/zenodo.6481141}.

\bibitem[{\c{C}}i{\c{c}}ek et~al.(2016){\c{C}}i{\c{c}}ek, Abdulkadir, Lienkamp,
  Brox, and Ronneberger]{cicek_3d_2016}
{\"O}zg{\"u}n {\c{C}}i{\c{c}}ek, Ahmed Abdulkadir, Soeren~S Lienkamp, Thomas
  Brox, and Olaf Ronneberger.
\newblock 3d u-net: learning dense volumetric segmentation from sparse
  annotation.
\newblock \emph{International conference on medical image computing and
  computer-assisted intervention}, pages 424--432, 2016.

\bibitem[Draulans et~al.(2021)Draulans, Pos, Smeenk, Kerkmeijer, Vogel,
  Nagarajah, Janssen, Mai, Heijmink, van~der Leest, et~al.]{draulans202168ga}
C{\'e}dric Draulans, Floris Pos, Robert~J Smeenk, Linda Kerkmeijer, Wouter~V
  Vogel, James Nagarajah, Marcel Janssen, Cindy Mai, Stijn Heijmink, Marloes
  van~der Leest, et~al.
\newblock 68ga-psma-11 pet, 18f-psma-1007 pet, and mri for gross tumor volume
  delineation in primary prostate cancer: intermodality and intertracer
  variability.
\newblock \emph{Practical Radiation Oncology}, 11\penalty0 (3):\penalty0
  202--211, 2021.

\bibitem[Eiber et~al.(2016)Eiber, Weirich, Holzapfel, Souvatzoglou, Haller,
  Rauscher, Beer, Wester, Gschwend, Schwaiger, et~al.]{eiber2016simultaneous}
Matthias Eiber, Gregor Weirich, Konstantin Holzapfel, Michael Souvatzoglou,
  Bernhard Haller, Isabel Rauscher, Ambros~J Beer, Hans-J{\"u}rgen Wester,
  Juergen Gschwend, Markus Schwaiger, et~al.
\newblock Simultaneous 68ga-psma hbed-cc pet/mri improves the localization of
  primary prostate cancer.
\newblock \emph{European urology}, 70\penalty0 (5):\penalty0 829--836, 2016.

\bibitem[Fedorov et~al.(2012)Fedorov, Beichel, Kalpathy-Cramer, Finet,
  Fillion-Robin, Pujol, Bauer, Jennings, Fennessy, Sonka, Buatti, Aylward,
  Miller, Pieper, and Kikinis]{slicer3dpaper}
Andriy Fedorov, Reinhard Beichel, Jayashree Kalpathy-Cramer, Julien Finet,
  Jean-Christophe Fillion-Robin, Sonia Pujol, Christian Bauer, Dominique
  Jennings, Fiona Fennessy, Milan Sonka, John Buatti, Stephen Aylward, James~V
  Miller, Steve Pieper, and Ron Kikinis.
\newblock {3D} slicer as an image computing platform for the quantitative
  imaging network.
\newblock \emph{Magnetic Resonance Imaging}, 30\penalty0 (9):\penalty0
  1323--1341, 2012.
\newblock URL \url{http://www.slicer.org}.

\bibitem[Gatidis et~al.(2023)Gatidis, Kustner, Ingrisch, Cyran, and
  Kleesiek]{gatidis_2023_autopet}
Sergios Gatidis, Thomas Kustner, Michael Ingrisch, Clemens Cyran, and Jens
  Kleesiek.
\newblock {Automated Lesion Segmentation in Whole-Body FDG- PET/CT - Domain
  Generalization}, April 2023.
\newblock URL \url{https://doi.org/10.5281/zenodo.7845727}.

\bibitem[Hadlich et~al.(2023)Hadlich, Marinov, and
  Stiefelhagen]{hadlich2023autopet}
Matthias Hadlich, Zdravko Marinov, and Rainer Stiefelhagen.
\newblock Autopet challenge 2023: Sliding window-based optimization of u-net.
\newblock \emph{arXiv preprint arXiv:2309.12114}, 2023.

\bibitem[Holzschuh et~al.(2023)Holzschuh, Mix, Ruf, H{\"o}lscher, Kotzerke,
  Vrachimis, Doolan, Ilhan, Marinescu, Spohn, Fechter, Kuhn, Bronsert, Gratzke,
  Grosu, Kamran, Heidari, Ng, K{\"o}nik, Grosu, and Zamboglou]{Holzschuh2023}
Julius~C. Holzschuh, Michael Mix, Juri Ruf, Tobias H{\"o}lscher, J{\"o}rg
  Kotzerke, Alexis Vrachimis, Paul Doolan, Harun Ilhan, Ioana~M. Marinescu,
  Simon~K.B. Spohn, Tobias Fechter, Dejan Kuhn, Peter Bronsert, Christian
  Gratzke, Radu Grosu, Sophia~C. Kamran, Pedram Heidari, Thomas~S.C. Ng, Arda
  K{\"o}nik, Anca-Ligia Grosu, and Constantinos Zamboglou.
\newblock Deep learning based automated delineation of the intraprostatic gross
  tumour volume in psma-pet for patients with primary prostate cancer.
\newblock \emph{Radiotherapy and Oncology}, 188, Nov 2023.

\bibitem[Isensee and Maier-Hein(2023)]{isensee2023look}
Fabian Isensee and Klaus~H Maier-Hein.
\newblock Look ma, no code: fine tuning nnu-net for the autopet ii challenge by
  only adjusting its json plans.
\newblock \emph{arXiv preprint arXiv:2309.13747}, 2023.

\bibitem[Isensee et~al.(2021)Isensee, Jaeger, Kohl, Petersen, and
  Maier-Hein]{nnU_Net_2020}
Fabian Isensee, Paul~F Jaeger, Simon A~A Kohl, Jens Petersen, and Klaus~H
  Maier-Hein.
\newblock nnu-net: a self-configuring method for deep learning-based biomedical
  image segmentation.
\newblock \emph{Nature methods}, 18\penalty0 (2):\penalty0 203—211, 2 2021.

\bibitem[Kerkmeijer et~al.(2021)Kerkmeijer, Groen, Pos, Haustermans,
  Monninkhof, Smeenk, Kunze-Busch, den Boer, Zijp, Vulpen,
  et~al.]{kerkmeijer2021focal}
Linda~GW Kerkmeijer, Veerle~H Groen, Floris~J Pos, Karin Haustermans, Evelyn~M
  Monninkhof, Robert~Jan Smeenk, MC~Kunze-Busch, JC~den Boer, JVDVV Zijp, M~van
  Vulpen, et~al.
\newblock Focal boost to the intraprostatic tumor in external beam radiotherapy
  for patients with localized prostate cancer: results from the flame
  randomized phase iii trial.
\newblock \emph{Journal of Clinical Oncology}, 39\penalty0 (7):\penalty0
  787--796, January 2021.

\bibitem[Kostyszyn et~al.(2020)Kostyszyn, Fechter, Bartl, Grosu, Gratzke,
  Sigle, Mix, Ruf, Fassbender, Kiefer, Bettermann, Nicolay, Spohn, Kramer,
  Bronsert, Guo, Qiu, Wang, Henkenberens, Werner, Baltas, Meyer, Derlin, Chen,
  and Zamboglou]{kostyszyn_intraprostatic_2020}
Dejan Kostyszyn, Tobias Fechter, Nico Bartl, Anca~L. Grosu, Christian Gratzke,
  August Sigle, Michael Mix, Juri Ruf, Thomas~F. Fassbender, Selina Kiefer,
  Alisa~S. Bettermann, Nils~H. Nicolay, Simon Spohn, Maria~U. Kramer, Peter
  Bronsert, Hongqian Guo, Xuefeng Qiu, Feng Wang, Christoph Henkenberens,
  Rudolf~A. Werner, Dimos Baltas, Philipp~T. Meyer, Thorsten Derlin, Mengxia
  Chen, and Constantinos Zamboglou.
\newblock Intraprostatic tumour segmentation on {PSMA}-{PET} images in patients
  with primary prostate cancer with a convolutional neural network.
\newblock \emph{Journal of Nuclear Medicine}, 2020.

\bibitem[Litjens et~al.(2017)Litjens, Kooi, Bejnordi, Setio, Ciompi,
  Ghafoorian, {van der Laak}, {van Ginneken}, and Sánchez]{litjens_2017}
Geert Litjens, Thijs Kooi, Babak~Ehteshami Bejnordi, Arnaud Arindra~Adiyoso
  Setio, Francesco Ciompi, Mohsen Ghafoorian, Jeroen~A.W.M. {van der Laak},
  Bram {van Ginneken}, and Clara~I. Sánchez.
\newblock A survey on deep learning in medical image analysis.
\newblock \emph{Medical Image Analysis}, 42:\penalty0 60--88, 2017.

\bibitem[Maier-Hein et~al.(2022)Maier-Hein, Reinke, Christodoulou, Glocker,
  Godau, Isensee, Kleesiek, Kozubek, Reyes, Riegler, et~al.]{maier2022metrics}
Lena Maier-Hein, Annika Reinke, Evangelia Christodoulou, Ben Glocker, Patrick
  Godau, Fabian Isensee, Jens Kleesiek, Michal Kozubek, Mauricio Reyes,
  Michael~A Riegler, et~al.
\newblock Metrics reloaded: Pitfalls and recommendations for image analysis
  validation.
\newblock \emph{arXiv preprint arXiv:2206.01653}, 2022.

\bibitem[Marhold et~al.(2022)Marhold, Kramer, Krainer, and {Le
  Magnen}]{MARHOLD2022304}
Maximilian Marhold, Gero Kramer, Michael Krainer, and Clémentine {Le Magnen}.
\newblock The prostate cancer landscape in europe: Current challenges, future
  opportunities.
\newblock \emph{Cancer Letters}, 526:\penalty0 304--310, 2022.

\bibitem[Milletari et~al.(2016)Milletari, Navab, and Ahmadi]{vnet_2016}
Fausto Milletari, Nassir Navab, and Seyed-Ahmad Ahmadi.
\newblock V-net: Fully convolutional neural networks for volumetric medical
  image segmentation.
\newblock \emph{2016 Fourth International Conference on 3D Vision (3DV)}, pages
  565--571, 2016.

\bibitem[Murugesan et~al.(2023)Murugesan, McCrumb, Brunner, Kumar, Soni,
  Grigorash, Moore, and Van~Oss]{murugesan2023autopet}
Gowtham~Krishnan Murugesan, Diana McCrumb, Eric Brunner, Jithendra Kumar, Rahul
  Soni, Vasily Grigorash, Stephen Moore, and Jeff Van~Oss.
\newblock Improving lesion segmentation in fdg-18 whole-body pet/ct scans using
  multilabel approach: Autopet ii challenge.
\newblock \emph{arXiv preprint arXiv:2311.01574}, 2023.

\bibitem[Myronenko(2019)]{segresnet2018}
Andriy Myronenko.
\newblock 3d mri brain tumor segmentation using autoencoder regularization.
\newblock \emph{Brainlesion: Glioma, Multiple Sclerosis, Stroke and Traumatic
  Brain Injuries}, pages 311--320, 2019.

\bibitem[Oktay et~al.(2018)Oktay, Schlemper, Folgoc, Lee, Heinrich, Misawa,
  Mori, McDonagh, Hammerla, Kainz, Glocker, and Rueckert]{attention_unet_2018}
Ozan Oktay, Jo~Schlemper, Lo{\"{\i}}c~Le Folgoc, Matthew C.~H. Lee, Mattias~P.
  Heinrich, Kazunari Misawa, Kensaku Mori, Steven~G. McDonagh, Nils~Y.
  Hammerla, Bernhard Kainz, Ben Glocker, and Daniel Rueckert.
\newblock Attention u-net: Learning where to look for the pancreas.
\newblock \emph{Medical Imaging with Deep Learning}, 2018.

\bibitem[Paszke et~al.(2019)Paszke, Gross, Massa, Lerer, Bradbury, Chanan,
  Killeen, Lin, Gimelshein, Antiga, et~al.]{pytorch_2019_NIPS}
Adam Paszke, Sam Gross, Francisco Massa, Adam Lerer, James Bradbury, Gregory
  Chanan, Trevor Killeen, Zeming Lin, Natalia Gimelshein, Luca Antiga, et~al.
\newblock Pytorch: An imperative style, high-performance deep learning library.
\newblock \emph{Advances in neural information processing systems}, 32, 2019.

\bibitem[Punn and Agarwal(2022)]{Punn2022}
Narinder~Singh Punn and Sonali Agarwal.
\newblock Modality specific u-net variants for biomedical image segmentation: a
  survey.
\newblock \emph{Artificial Intelligence Review}, Mar 2022.

\bibitem[Ronneberger et~al.(2015)Ronneberger, Fischer, and
  Brox]{ronneberger_u_net_2015}
Olaf Ronneberger, Philipp Fischer, and Thomas Brox.
\newblock U-net: Convolutional networks for biomedical image segmentation.
\newblock \emph{Medical Image Computing and Computer-Assisted Intervention –
  {MICCAI} 2015}, pages 234--241, 2015.

\bibitem[Saha et~al.(2022)Saha, Twilt, Bosma, van Ginneken, Yakar, Elschot,
  Veltman, Fütterer, de~Rooij, and Huisman]{saha_anindo_2022_6522364}
Anindo Saha, Jasper~Jonathan Twilt, Joeran~Sander Bosma, Bram van Ginneken,
  Derya Yakar, Mattijs Elschot, Jeroen Veltman, Jurgen Fütterer, Maarten
  de~Rooij, and Henkjan Huisman.
\newblock {Artificial Intelligence and Radiologists at Prostate Cancer
  Detection in MRI: The PI-CAI Challenge (Study Protocol)}, May 2022.
\newblock URL \url{https://pi-cai.grand-challenge.org}.

\bibitem[Siegel et~al.(2024)Siegel, Giaquinto, and
  Jemal]{siegel_2024_CancerStatistics}
Rebecca~L. Siegel, Angela~N. Giaquinto, and Ahmedin Jemal.
\newblock Cancer statistics, 2024.
\newblock \emph{CA: A Cancer Journal for Clinicians}, 74\penalty0 (1):\penalty0
  12--49, 2024.

\bibitem[Spohn et~al.(2020)Spohn, Kramer, Kiefer, Bronsert, Sigle,
  Schultze-Seemann, Jilg, Sprave, Ceci, Fassbender,
  et~al.]{spohn2020comparison}
Simon~KB Spohn, Maria Kramer, Selina Kiefer, Peter Bronsert, August Sigle,
  Wolfgang Schultze-Seemann, Cordula~A Jilg, Tanja Sprave, Lara Ceci, Thomas~F
  Fassbender, et~al.
\newblock Comparison of manual and semi-automatic [18f] psma-1007 pet based
  contouring techniques for intraprostatic tumor delineation in patients with
  primary prostate cancer and validation with histopathology as standard of
  reference.
\newblock \emph{Frontiers in Oncology}, 10:\penalty0 600690, 2020.

\bibitem[Sung et~al.(2021)Sung, Ferlay, Siegel, Laversanne, Soerjomataram,
  Jemal, and Bray]{sung2021global}
Hyuna Sung, Jacques Ferlay, Rebecca~L Siegel, Mathieu Laversanne, Isabelle
  Soerjomataram, Ahmedin Jemal, and Freddie Bray.
\newblock Global cancer statistics 2020: Globocan estimates of incidence and
  mortality worldwide for 36 cancers in 185 countries.
\newblock \emph{CA: a cancer journal for clinicians}, 71\penalty0 (3):\penalty0
  209--249, 2021.

\bibitem[Tamal(2020)]{TAMAL2020e05267}
Mahbubunnabi Tamal.
\newblock Intensity threshold based solid tumour segmentation method for
  positron emission tomography (pet) images: A review.
\newblock \emph{Heliyon}, 6\penalty0 (10):\penalty0 e05267, 2020.

\bibitem[Xu et~al.(2023)Xu, Klyuzhin, Harsini, Ortiz, Zhang, Bénard, Dodhia,
  Uribe, Rahmim, and Lavista~Ferres]{xu_automatic_2023}
Yixi Xu, Ivan Klyuzhin, Sara Harsini, Anthony Ortiz, Shun Zhang, François
  Bénard, Rahul Dodhia, Carlos~F. Uribe, Arman Rahmim, and Juan
  Lavista~Ferres.
\newblock Automatic segmentation of prostate cancer metastases in {PSMA}
  {PET}/{CT} images using deep neural networks with weighted batch-wise dice
  loss.
\newblock \emph{Computers in Biology and Medicine}, 158:\penalty0 106882, May
  2023.

\bibitem[Zamboglou et~al.(2019)Zamboglou, Fassbender, Steffan, Schiller,
  Fechter, Carles, Kiefer, Rischke, Reichel, Schmidt-Hegemann, Ilhan,
  Chirindel, Nicolas, Henkenberens, Derlin, Bronsert, Mavroidis, Chen, Meyer,
  Ruf, and Grosu]{Zamboglou2019}
Constantinos Zamboglou, Thomas~F. Fassbender, Lina Steffan, Florian Schiller,
  Tobias Fechter, Montserrat Carles, Selina Kiefer, Hans~C. Rischke, Kathrin
  Reichel, Nina-Sophie Schmidt-Hegemann, Harun Ilhan, Alin~F. Chirindel,
  Guillaume Nicolas, Christoph Henkenberens, Thorsten Derlin, Peter Bronsert,
  Panayiotis Mavroidis, Ronald~C. Chen, Philipp~T. Meyer, Juri Ruf, and Anca~L.
  Grosu.
\newblock Validation of different psma-pet/ct-based contouring techniques for
  intraprostatic tumor definition using histopathology as standard of
  reference.
\newblock \emph{Radiotherapy and Oncology}, 141:\penalty0 208--213, Dec 2019.

\bibitem[Zamboglou et~al.(2022)Zamboglou, Spohn, Ruf, Benndorf, Gainey, Kamps,
  Jilg, Gratzke, Adebahr, Schmidtmayer-Zamboglou, et~al.]{zamboglou2022psma}
Constantinos Zamboglou, Simon~KB Spohn, Juri Ruf, Matthias Benndorf, Mark
  Gainey, Marius Kamps, Cordula Jilg, Christian Gratzke, Sonja Adebahr, Barbara
  Schmidtmayer-Zamboglou, et~al.
\newblock Psma-pet-and mri-based focal dose escalated radiation therapy of
  primary prostate cancer: Planned safety analysis of a nonrandomized 2-armed
  phase 2 trial (aro2020-01).
\newblock \emph{International Journal of Radiation Oncology* Biology* Physics},
  113\penalty0 (5):\penalty0 1025--1035, 2022.

\bibitem[Zhou et~al.(2019)Zhou, Siddiquee, Tajbakhsh, and Liang]{unet_pp_2018}
Zongwei Zhou, Md~Mahfuzur~Rahman Siddiquee, Nima Tajbakhsh, and Jianming Liang.
\newblock Unet++: Redesigning skip connections to exploit multiscale features
  in image segmentation.
\newblock \emph{IEEE transactions on medical imaging}, 39\penalty0
  (6):\penalty0 1856--1867, 2019.

\end{thebibliography}
\end{document}


\maketitle

\section{Appendix A - Dataset and Code}

\subsection{Dataset Collection}

All the datasets used in this paper were collected at our partner sites. Unfortunately, due to certain restrictions, these datasets will not be made available to the public.

\paragraph{68-Ga: }One hour after intravenous tracer injection, patients underwent a whole body PET scan after voiding. Whole body acquisition protocols were acquired on three different Philips scanners: GEMINI TF TOF 64, GEMINI TF 16 Big Bore and Vereos. All systems resulted in a PET image with a voxel size of 2 x 2 x 2 mm3. Images were normalized to decay corrected injected activity per kg body weight (standardized uptake values, SUV in [g/ml]).

\paragraph{18-F: }The mean injected activity of [18F]PSMA-1007 was 299 MBq (min–max: 249–370 MBq). Patients underwent a whole-body PET scan starting 2 hours after injection. Scans were performed with a 16-slice Gemini TF big bore and a Vereos PET/CT scanner. At the time of the PET scan, either a contrast-enhanced or native diagnostic CT scan (120 kVp, 100–400 mAs, dose modulation) was performed for attenuation correction. All systems resulted in a PET image with a voxel size of 2 × 2 × 2 mm3. 

\subsection{Ethics Statement}
The studies involving human participants were reviewed and approved by the Institutional Review Board at the partner sites. The physicians at the partner sites anonymised the images before sending to us. The medical images (PET, prostate, and tumour contours) do not contain any personally identifiable information or offensive content.

\subsection{Code}
The source code (modified nnU-Net framework~\cite{nnU_Net_2020}) and instructions to reproduce the experiments will be stored in a separate folder, and will be publicly available in a GitHub repository. Anyone who wishes to replicate these can simply follow the instructions provided.

\section{Appendix B - Results}

Tables~\ref{table:AllNormResults_DSC_Metrics} and~\ref{table:AllNormResults_HD95_Metrics} showcase the segmentation accuracies of all the four models for both datasets across four different normalisation methods. It is clearly evident that the networks that we preprocessed with FCN are superior to the existing normalisation schemes. We also conducted model-wise statistical significance tests (Wilcoxon-signed rank test) to compare the differences in performance between the top ranking normalisation scheme (FCN) versus the remaining three (Z-score, CT and fixed clip). This is shown in Table~\ref{table:Stats_Test}.

\begin{table*}[ht]
\fontsize{9}{10}\selectfont
\centering
\begin{tabular}{|c|c|c|c|c|c|c|c|}
\hline                  
PSMA & Training & Testing & Model & Z-score & CT & fixed clip: & FCN \\
tracer & data & data & name &~\cite{nnU_Net_2020} &~\cite{nnU_Net_2020} &$0$-$15$~\cite{Holzschuh2023} & (ours) \\ 
\hline
\multirow{4}{*}{68-Ga} & & & UNet & $ 0.755 $ & $ 0.763 $ & $ 0.770 $ & $ \mathbf{ 0.789 } $ \\ 
& Centres A + B & Centre A & IB-UNet & $ 0.765 $ & $ 0.770 $ & $ 0.779 $ & $ \mathbf{ 0.829 } $ \\ 
& \textit{n} = 151 & \textit{n} = 17 & Att. UNet & $ 0.747 $ & $ 0.774 $ & $ 0.775 $ & $ \mathbf{ 0.807 } $ \\ 
& & & IB-Att. UNet & $ 0.761 $ & $ 0.778 $ & $ 0.780 $ & $ \mathbf{ 0.831 } $ \\ 
\hline
\multirow{8}{*}{18-F} & & & UNet & $ 0.776 $ & $ 0.790 $ & $ 0.787 $ & $ \mathbf{ 0.816 } $ \\ 
& & Centre A & IB-UNet & $ 0.783 $ & $ 0.792 $ & $ 0.791 $ & $ \mathbf{ 0.833 } $ \\ 
& & \textit{n} = 19 & Att. UNet & $ 0.776 $ & $ 0.793 $ & $ 0.787 $ & $ \mathbf{ 0.816 } $ \\ 
& Centre A & & IB-Att. UNet & $ 0.781 $ & $ 0.800 $ & $ 0.793 $ & $ \mathbf{ 0.840 } $ \\ 
\cline{3-8}
& \textit{n} = 131 & & UNet & $ 0.615 $ & $ 0.657 $ & $ 0.661 $ & $ \mathbf{ 0.717 } $ \\ 
& & Centre C & IB-UNet & $ 0.630 $ & $ 0.659 $ & $ 0.665 $ & $ \mathbf{ 0.730 } $ \\ 
& & \textit{n} = 14 & Att. UNet & $ 0.625 $ & $ 0.662 $ & $ 0.662 $ & $ \mathbf{ 0.722 } $ \\ 
& & & IB-Att. UNet & $ 0.651 $ & $ 0.670 $ & $ 0.668 $ & $ \mathbf{ 0.747 } $ \\ 
\hline
\end{tabular}
\vspace{2mm}
\caption{Comparison of U-Net variants all implemented by us in the same nnU-Net framework, for different normalisation methods, using DSC metric ($\uparrow$).~\textit{Z-score} standardises voxels using local mean and standard deviation.~\textit{CT} first does percentile clipping and then standardises voxels using global mean and standard deviation.~\textit{fixed clip} limits intensities to pre-defined lower and upper bounds. Statistical significance difference between best performing FCN preprocessing scheme against the rest of the normalisations is in bold.}
\label{table:AllNormResults_DSC_Metrics}
\end{table*}

\begin{table*}[ht]
\fontsize{9}{10}\selectfont
\centering
\begin{tabular}{|c|c|c|c|c|c|c|c|}
\hline                  
PSMA & Training & Testing & Model & Z-score & CT & fixed clip: & FCN \\
tracer & data & data & name &~\cite{nnU_Net_2020} &~\cite{nnU_Net_2020} &$0$-$15$~\cite{Holzschuh2023} & (ours) \\ 
\hline
\multirow{4}{*}{68-Ga} & & & UNet & $ 7.978 $ & $ 7.209 $ & $ 6.491 $ & $ \mathbf{ 6.057 } $ \\ 
& Centres A + B & Centre A & IB-UNet & $ 7.029 $ & $ 6.491 $ & $ 5.673 $ & $ \mathbf{ 5.350 } $ \\ 
& \textit{n} = 151 & \textit{n} = 17 & Att. UNet & $ 8.584 $ & $ 6.158 $ & $ 6.106 $ & $ \mathbf{ 5.531 } $ \\ 
& & & IB-Att. UNet & $ 7.414 $ & $ 5.784 $ & $ 5.590 $ & $ \mathbf{ 4.917 } $ \\ 
\hline
\multirow{8}{*}{18-F} & & & UNet & $ 5.977 $ & $ 4.427 $ & $ 4.787 $ & $ \mathbf{ 3.691 } $ \\ 
& & Centre A & IB-UNet & $ 5.174 $ & $ 4.233 $ & $ 4.288 $ & $ \mathbf{ 3.518 } $ \\ 
& & \textit{n} = 19 & Att. UNet & $ 6.004 $ & $ 4.067 $ & $ 4.759 $ & $ \mathbf{ 3.499 } $ \\ 
& Centre A & & IB-Att. UNet & $ 5.424 $ & $ 3.408 $ & $ 5.202 $ & $ \mathbf{ 3.354 } $ \\ 
\cline{3-8}
& \textit{n} = 131 & & UNet & $ 8.757 $ & $ 7.773 $ & $ 7.414 $ & $ \mathbf{ 5.108 } $ \\ 
& & Centre C & IB-UNet & $ 7.491 $ & $ 7.542 $ & $ 7.029 $ & $ \mathbf{ 4.868 } $ \\ 
& & \textit{n} = 14 & Att. UNet & $ 8.800 $ & $ 7.311 $ & $ 7.234 $ & $ \mathbf{ 4.948 } $ \\ 
& & & IB-Att. UNet & $ 7.363 $ & $ 7.240 $ & $ 6.747 $ & $ \mathbf{ 4.332 } $ \\ 
\hline
\end{tabular}
\vspace{2mm}
\caption{Comparison of U-Net variants all implemented by us in the same nnU-Net framework, for different normalisation methods, using HD-95 metric ($\downarrow$).}
\label{table:AllNormResults_HD95_Metrics}
\end{table*}

\begin{table*}[ht]
\fontsize{9}{10}\selectfont
\centering
\begin{tabular}{|c|c|c|c|c|c|}
\hline                  
Tracer & Model & Normalisation & NSD & HD-95 & DSC\\\
Type & Name & Schemes &~\textit{p-value} &~\textit{p-value} &~\textit{p-value} \\
\hline
\multirow{16}{*}{68-Ga} & \multirow{4}{*}{U-Net} & FCN \textit{vs} Z-score & $0.0055$ & $0.0023$ & $0.0093$\\
& & FCN \textit{vs} CT &  $0.0174$ & $0.0149$ & $0.0154$ \\
& & FCN \textit{vs} fixed clip &$0.0311$ & $0.0298$ & $0.0325$  \\
\cline{2-6}
& \multirow{4}{*}{IB-U-Net} & FCN \textit{vs} Z-score & $0.0057$ & $0.0036$ & $0.0009$  \\
& & FCN \textit{vs} CT & $0.0254$ & $0.0244$ & $0.0260$ \\
& & FCN \textit{vs} fixed clip & $0.0293$ & $0.0287$ & $0.0333$ \\
\cline{2-6}
& \multirow{4}{*}{Att. U-Net} & FCN \textit{vs} Z-score & $0.0011$ & $0.0030$ & $0.0055$ \\
& & FCN \textit{vs} CT & $0.0196$ & $0.0223$ & $0.0259$ \\
& & FCN \textit{vs} fixed clip & $0.0194$ & $0.0212$ & $0.0236$ \\
\cline{2-6}
& \multirow{4}{*}{IB-Att. U-Net} & FCN \textit{vs} Z-score & $0.0041$ & $0.0034$ & $0.0016$ \\
& & FCN \textit{vs} CT & $0.0156$ & $0.0148$ & $0.0111$ \\
& & FCN \textit{vs} fixed clip & $0.0251$ & $0.0274$ & $0.0201$ \\
\hline
\multirow{16}{*}{18-F} & \multirow{4}{*}{U-Net} & FCN \textit{vs} Z-score & $0.0087$ & $0.0064$ & $0.0044$ \\
& & FCN \textit{vs} CT & $0.0275$ & $0.0242$ & $0.0268$ \\
& & FCN \textit{vs} fixed clip & $0.0154$ & $0.0166$ & $0.0143$ \\
\cline{2-6}
& \multirow{4}{*}{IB-U-Net} & FCN \textit{vs} Z-score & $0.0068$ & $0.0081$ & $0.0031$ \\
& & FCN \textit{vs} CT & $0.0393$ & $0.0338$ & $0.0309$ \\
& & FCN \textit{vs} fixed clip & $0.0204$ & $0.0161$ & $0.0150$ \\
\cline{2-6}
& \multirow{4}{*}{Att. U-Net} & FCN \textit{vs} Z-score & $0.0084$ & $0.0007$ & $0.0040$ \\
& & FCN \textit{vs} CT & $0.0280$ & $0.0296$ & $0.0342$ \\
& & FCN \textit{vs} fixed clip & $0.0101$ & $0.0043$ & $0.0089$  \\
\cline{2-6}
& \multirow{4}{*}{IB-Att. U-Net} & FCN \textit{vs} Z-score & $0.0077$ & $0.0042$ & $0.0001$\\
& & FCN \textit{vs} CT & $0.0454$ & $0.0499$ & $0.0396$ \\
& & FCN \textit{vs} fixed clip & $0.0138$ & $0.0111$ & $0.0120$  \\
\hline
\end{tabular}
\vspace*{2mm}
\caption{Statistical comparison of the U-Net variants based on best performing normalisation scheme against the rest of the normalisations. These values are based on testing results for both 68-Ga and 18-F datasets from Centre A, and there is significant difference if $p\,{<}\,0.05$.}
\label{table:Stats_Test}
\end{table*}

\bibliography{references}